\documentclass[aps,prb,twocolumn,prabib,amsmath,amssymb,superscriptaddress,floatfix]{revtex4-1}
\usepackage{graphics}
\usepackage{bm}
\usepackage{dcolumn}
\usepackage{natbib}
\usepackage{epstopdf}
\usepackage{graphicx}
\usepackage{epsfig}
\usepackage{color,xcolor}
\usepackage[pdfstartview=FitH,colorlinks,bookmarks=false,citecolor=blue,linkcolor=red,urlcolor=blue]{hyperref}
\newcommand{\kk}{{\bf k}}

\newcommand{\qq}{{\bf q}}
\newcommand{\rr}{{\bf r}}
\begin{document}
\title{Bosonic integer quantum Hall states in topological bands with Chern number two}
\author{Tian-Sheng Zeng}
\affiliation{Department of Physics and Astronomy, California State University, Northridge, California 91330, USA}
\author{W. Zhu}
\affiliation{Department of Physics and Astronomy, California State University, Northridge, California 91330, USA}
\author{D. N. Sheng}
\affiliation{Department of Physics and Astronomy, California State University, Northridge, California 91330, USA}
\date{\today}
\begin{abstract}
  We study the interacting bosons in topological Hofstadter bands with Chern number two. Using exact diagonalization, we demonstrate that bosonic integer quantum Hall (BIQH) state emerges at integer boson filling factor $\nu=1$ of the lowest Chern band with evidences including a robust spectrum gap and quantized topological Hall conductance two. Moreover, the robustness of BIQH state against different interactions and next-nearest neighbor hopping is investigated. The strong nearest neighbor interaction would favor a charge density wave. When the onsite interaction decreases, BIQH state undergoes a continuous transition  into a superfluid state. Without next-nearest neighbor hopping, the ground state is possibly in  a metallic Fermi-liquid-like phase.
\end{abstract}
\maketitle
\section{Introduction}
Recent theoretical studies reveal that in two dimensions strongly interacting two-component bosons in a magnetic field can realize a bosonic integer quantum Hall (BIQH) state~\cite{Senthil2013,Furukawa2013,Regnault2013,Wu2013,Grass2014}. The BIQH phase characterized by Hall conductivity quantized to an even integer~\cite{Lu2012,Wen2013,Geraedts2013,Mulligan2014} is protected by a global $U(1)$-symmetry and the real-space entanglement spectrum of this state hosts two counter propagating chiral modes. Recently, two different lattice versions of BIQH states have been proposed at integer filling $\nu=1$ of the lowest topological flat-band with Chern number $C=2$. The optical flux lattice has been studied by exact diagonalization of the projected Hamiltonian in momentum space~\cite{Sterdyniak2015} and the correlated Haldane-honeycomb lattice has been studied by infinite density matrix renormalization group of hardcore boson in real space~\cite{He2015}. This is different from the two dimensional topological $C=1$ band filled by hardcore bosons at $\nu=1$, which  is believed to exhibit the Fermi-liquid-like state~\cite{Read1998}.
Indeed, for $C>1$, a series of color-entangled Abelian topological states have been suggested at various filling numbers under repulsive two-body interaction~\cite{LBFL2012,Wang2012r,Yang2012,Sterdyniak2013,Wang2013}. In Harper-Hofstadter model with topological $C=2$ band, different emergent topological  states including the symmetry protected BIQH state, can be understood by an insightful approach from Streda formula of composite fermion~\cite{Cooper2015}.
So far, the study on BIQH state for single component bosons on topological Hofstadter lattice is still lacking. It is interesting to compare the BIQH in such a system with other lattice realizations of topological flatbands with $C=2$.  More specifically,
it is interesting to address the issue what conditions can make the BIQH stable against possible competing phases, like charge density wave and bosonic superfluid. If the system can host other competing phases under certain conditions, it opens a door to explore quantum phase transition between BIQH and other phases. In Refs.~\cite{Grover2013,Lu2014}, the low energy theory describing continuous phase transitions between superfluid and BIQH is constructed from fermionic parton.

In this paper, we study the generalized Hofstadter model and address the stability of the BIQH against other phases, taking into account the effects of interaction strength, band topology and their interplay within the full real space Hamiltonian. We find that the many-body ground state is indeed BIQH for hardcore bosons at integer filling of $C=2$ band with a robust spectrum gap, due to onsite Hubbard repulsion. Without onsite Hubbard repulsion, the softcore bosons would undergo a Bose-condensation into the lowest single particle orbit. Increasing the nearest neighbor interaction to strong repulsion, a charge density wave state would dominate over the BIQH state. When tuning the next nearest neighbor hopping down to zero, the Chern number of
the lowest band becomes $C=-1$ and we show that the ground state would have a transition into a metallic liquid-like phase in this case. Thus, the emergence of BIQH phase in single component bosons on lattice model is ultimately related to the interplay of interaction and band topology. Experimentally, the bosonic Hofstadter model has been realized by laser-assisted tunneling in cold atoms~\cite{Aidelsburger2013,Miyake2013}, and a Bose-Einstein condensation (BEC) is observed in this cold atom setup~\cite{Kennedy2015}. In relation to current experiments, we discuss an experimental prospect toward realization of BIQH state in optical lattices and further predict that once a stronger next nearest neighbor hopping can be implemented in this setup, one may realize the many-body BIQH state in such a generalized Hofstadter model and study the quantum phase transition between Bose condensate and BIQH phase by tuning the onsite interaction between bosons through lattice potential or Feshbach resonances~\cite{Courteille1998}.

This paper is organized as follows. In Sec.~\ref{model}, we give a description of the Bose-Hubbard model on the generalized Hofstadter band with topological invariant $C=2$. In Sec.~\ref{mbgs}, we explore the many-body ground state at infinite onsite repulsion (namely, hardcore boson) and present a detailed proof of BIQH state by exact diagonalization at filling $\nu=1$. In Sec.~\ref{phasetransition}, we explore other possible competing phases like bosonic superfluid, and discuss their transitions into BIQH state by varying
interactions and lattice parameters. Finally, in Sec.~\ref{summary}, we summarize our results and discuss the prospect of investigating nontrivial topological states in cold atoms systems.

\section{The Bose-Hubbard Model}\label{model}
Here, we consider the interacting bosons on the generalized Hofstadter lattice which describes the motion of charged particle under a uniform magnetic field $B$ on a square lattice with lattice constant $a=1$~\cite{Hofstadter1976}. Take the gauge vector $A=B(\frac{1}{2},x,0)$, such that the particle hopping between sites $\rr$ and $\rr'$ has a magnetic phase $\exp(iA\cdot(\rr-\rr'))$~\cite{Kapit2010}. In cold atoms, this spatially dependent gauge field can be artificially engineered in laser-assisted tunneling. When both nearest neighbor and next nearest neighbor hoppings are included, the noninteracting generalized Hofstadter model, is described by the Hamiltonian~\cite{Wang2013}
\begin{align}
  H_0=&-\sum_{\rr}t_xb^{\dag}_{\rr}b_{\rr+i_x}+t_ye^{i\phi x}b^{\dag}_{\rr}b_{\rr+i_y}+h.c.\nonumber\\
  &-t'\sum_{\rr}e^{i\phi(x\pm\frac{1}{2})}b^{\dag}_{\rr}b_{\rr\pm i_x+i_y}+h.c.
\end{align}
where $b_{\rr}$ is the bosonic annihilating operator at site $\rr=(x,y)$, $i_{\alpha}$ the unit vector along the $\alpha$-direction and the magnetic flux through each plaquette $\phi=Ba^2$. Now we consider that the bosons interact with each other via:
\begin{align}
  V_{int}=\frac{U}{2}\sum_{\rr} n_{\rr}(n_{\rr}-1)+V\sum_{\langle\rr,\rr'\rangle} n_{\rr}n_{\rr'},
\end{align}
where $n_{\rr}$ is the boson number operator and $\langle\rr,\rr'\rangle$ denote nearest neighbor pairs of sites. Here we take lattice parameters $\phi/2\pi=1/3,t_x=t_y=t,t'=-0.5t$, such that the lowest band has a topological invariant $C=2$. When $t'=0$, the lowest band has a topological invariant $C=-1$. We can choose three sites in the x direction as a magnetic unit cell and the total number of lattice sites is $N_L=3N_s$, $N_s=N_x\times N_y$ is the number of unit cells. We explore the many-body ground state of $H=H_0+V_{int}$ by exactly diagonalizing a finite $N$-particle system at fixed integer filling $\nu=N/N_s=1$. With periodic conditions we identify each many-body state using its total momentum sectors $(2\pi k_x/N_x, 2\pi k_y/N_y)$ due to translation symmetry.

\section{Bosonic Integer Quantum Hall state}\label{mbgs}

\begin{figure}[t]
  \includegraphics[height=3.0in,width=3.4in]{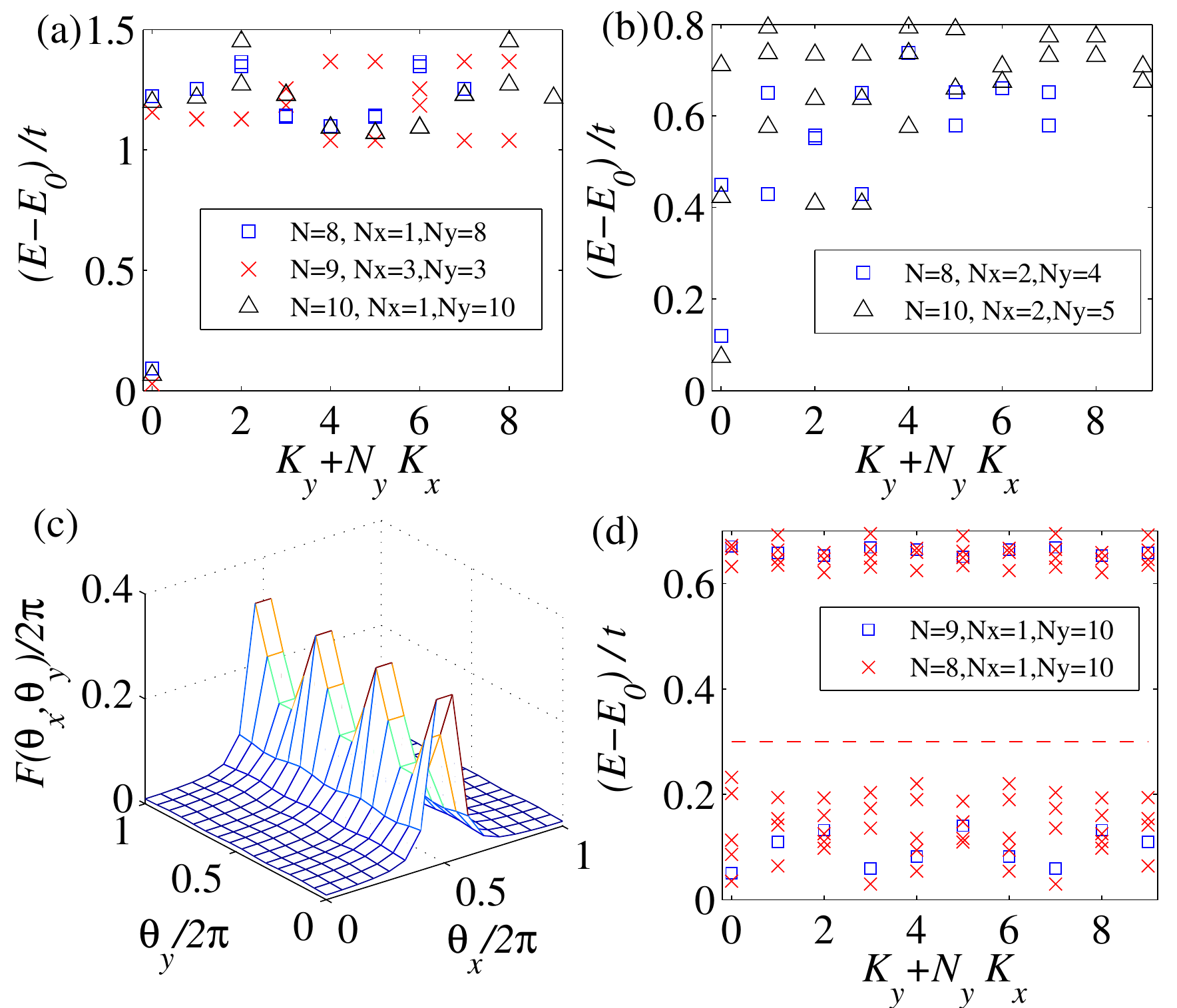}
  \caption{\label{energy}(Color online) Numerical results for generalized Hofstadter model with infinite two-body interaction $U=\infty$: (a-b) Low energy spectrum of different system sizes at filling $\nu=1$; (c) The Berry curvatures for the $K=(0,0)$ ground state of $N=8,N_x=2,N_y=4,t'=-0.5$ system at $16\times16$ mesh points; (d) Low energy spectrum of one or two quasiholes by removing particles. The number of quasihole energy manifold under the red dashed line matches that of BIQH state.}
\end{figure}

We first look at the limiting case that $U=\infty$ and $V=0$, where each site can be occupied by one boson at most. The low energy spectrum is plotted in Figs.~\ref{energy}(a-b), for systems with different aspect ratios $N_y/3N_x$. We find that there always exists a single gapped ground state with total momentum $K=(K_x,K_y)=(0,0)$, separated from the excited states by a large gap, for both even and odd $N_s$.

The BIQH state is characterized by a finite spectrum gap. To explore the stability of this phase, we use the twisted boundary conditions $\psi(\rr+N_{\alpha})=e^{i\theta_{\alpha}}\psi(\rr),\alpha=x,y$, and inspect the spectrum gap. The energy gap $\Delta$ is defined as the minimum value of the difference of the ground energy and the first excited energy on the full $(\theta_x,\theta_y)$ parameter plane in order to explore the stability of the phase. If the ground state mixes with other levels
during the change of the boundary phases, we take $\Delta=0$. Our calculations confirm that the obtained
 ground state does not mix with the higher energy levels, demonstrating itself as the unique ground state under the insertion of
the flux  $\theta_{\alpha}$. The many-body Chern number of the ground state wavefunction $\psi$ at $K=(0,0)$ is given by
\begin{align}
  C_{mb}=\frac{1}{2\pi}\int_{0}^{2\pi}d\theta_x\int_{0}^{2\pi}d\theta_yF(\theta_x,\theta_y),
\end{align}
where  $F(\theta_x,\theta_y)=\mathbf{Im}(\langle{\partial_{\theta_x}\psi}|{\partial_{\theta_y}\psi}\rangle
-\langle{\partial_{\theta_y}\psi}{\partial_{\theta_x}\psi}\rangle)$ is the Berry curvature. By numerically calculating the Berry curvatures using $m\times m$ mesh points in boundary phase space with $m\geq9$ as shown in Fig.~\ref{energy}(c), we find that the many-body Chern number $C_{mb}$ indeed converges to a quantized value 2. In addition, we calculate the density structure factor
\begin{align}
  S(\qq)=\frac{1}{N_{L}}\sum_{\rr,\rr'}e^{i\qq\cdot(\rr-\rr')}\left(\langle n_{\rr}n_{\rr'}\rangle-\langle n_{\rr}\rangle\langle n_{\rr'}\rangle\delta_{\qq,0}\right)
\end{align}
for the ground state, and we find no evidence of finite momentum Bragg peaks in $S(\qq)$. Thus we can rule out the possibility of charge density wave (CDW) as the competing ground state.

To distinguish the BIQH phase from the usual topological ordered phase, we investigate the quasiparticle excitations in the bulk. In a topological ordered phase such as fractional quantum Hall states, the existence of quasihole excitations carrying fractional charge is the key evidence for the nontrivial nature of the topological state. Here, we utilize two different methods. First, we generate a quasihole by inserting a single flux quantum, namely we change the flux quanta of the lattice to $N_s=N+1$. Here as shown in Fig.~\ref{energy}(d), the counting number of low energy states of $N$ particles in $N_s$ orbits is simply given by  $N_s\text{!}/(N_s-N)\text{!}/N\text{!}$. Second, by introducing an onste impurity potential $V_{imp}=\sum_{\rr}\delta_{\rr,\rr_0} n_{\rr}$ at site $\rr_0$ of a given unit cell~\cite{Liu2015}, one can pin the quasihole near the impurity location, and define the excess charge $Q=\sum_{|\rr-\rr_0|\leq2}(n_{\rr}-\bar{n})$ as the quasihole charge, where $\bar{n}$ is the uniform density of BIQH state. For example, we add a $\delta$-impurity potential to the original periodic system $N=9,N_x=1,N_y=10$ and obtain $Q=1$ to high precision, thus we obtain the exciations carrying integer charge unit, in against to the fracional value as expected for topological ordered phases. These numerical results match the theoretical predictions of BIQH state without any bulk topological order.

\section{Competing phases and Phase Transitions}\label{phasetransition}

\begin{figure}[t]
  \includegraphics[height=2.8in,width=3.4in]{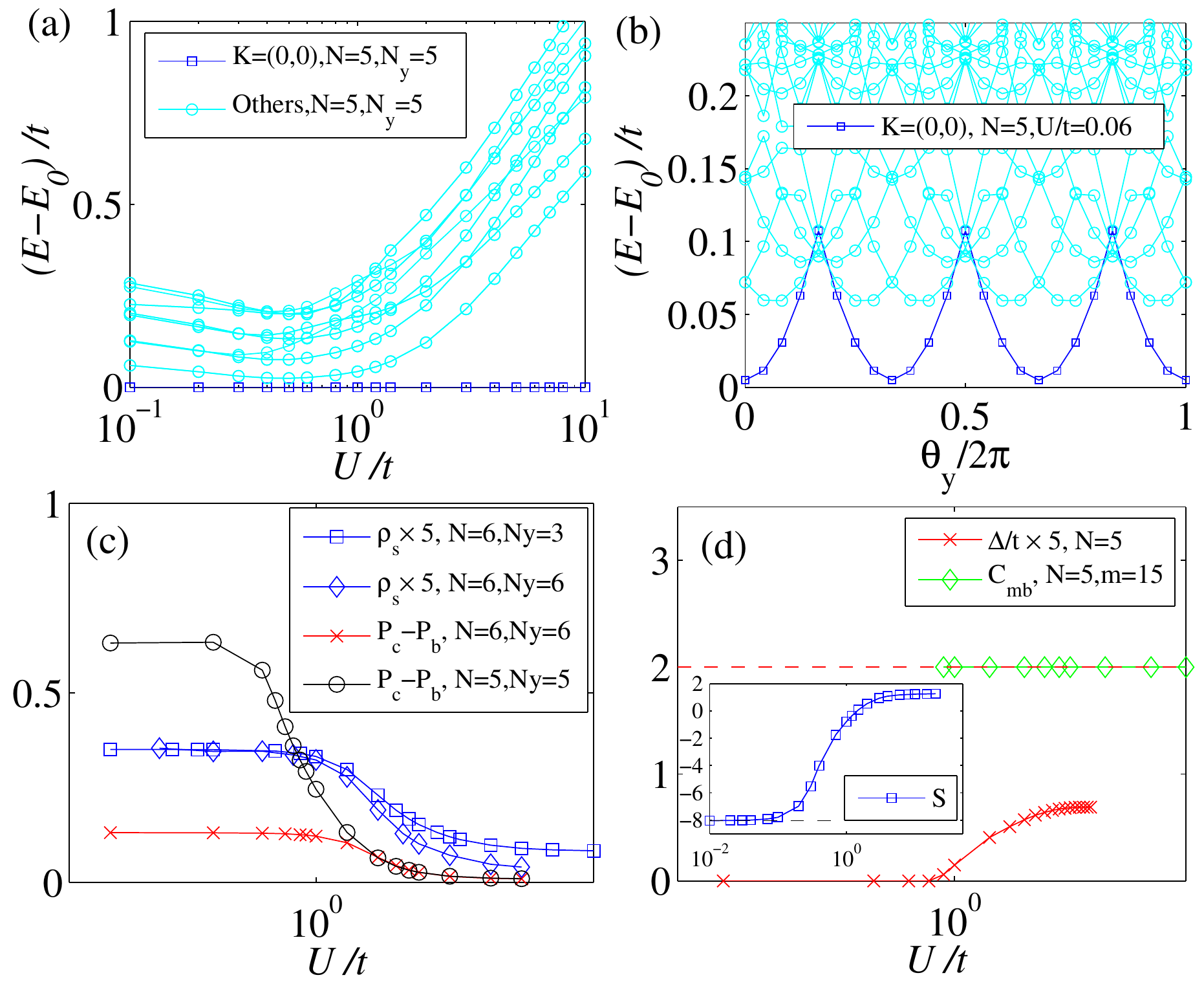}
  \caption{\label{phaseU}(Color online) Numerical results for generalized Hofstadter model with softcore two-body interaction $U$: (a) the evolutions of the energy difference between the $K=(0,0)$ ground energy and the excited levels at $(\theta_x,\theta_y)=(0,0)$; (b) The low energy spectra  under the insertion of $\theta_y$ at fixed $U/t=0.06$ and $\theta_x=0$; (c) the diffraction peak of the visibility $P_c-P_b$ and the related superfluid phase stiffness $\rho_s$; (d) The energy gap $\Delta$ and many-body Chern number of the groundstate  at $K=(0,0)$ calculated using  $9\times9$ and $15\times15$ mesh points in boundary phase space for Berry curvature. The red dashed line in (d) panel is the quantized value $C=2$. The inset panel shows the one-particle occupancy entropy.}
\end{figure}

Having established BIQH phase in the condition of $U=\infty$ and $V=0$, we continue to
discuss the competing phases in the nearby parameter space by varying interaction strength $U$, $V$ and
the band parameter $t'$, respectively. The corresponding phase transitions will also be addressed.

\subsection{Superfluid phase}

Following the last section, we further consider the effect of a finite repulsion $U$ and small $t'$ on possible competing phases with zero nearest interaction $V=0$. As it is well-known, without onsite interaction, the free bosons would undergo a condensation into the lowest single-particle energy orbit for unfrustrated energy bands. The emergence of BIQH state in the strongly interacting regime signifies the important role of the band topology and interaction. In Fig.~\ref{phaseU}(a), we plot the energy variation of $N=5,N_y=5,t'=-0.5t$ against $U$ at $(\theta_x,\theta_y)=(0,0)$. The $K=(0,0)$ ground state evolves adiabatically with $U$. At small $U/t\ll1$, the low energy spectrum is shown in Fig.~\ref{phaseU}(b), and the $K=(0,0)$ ground state mixes with excited levels under the insertion of $\theta_y$, demonstrating its metallic (gapless) nature.

On one hand, Bose condensate can be characterized by the off-diagonal long range order $\rho_{\rr,\rr'}=\langle\psi|b_{\rr}^{\dag}b_{\rr'}|\psi\rangle$, such that the condensation  momentum  can be identified by the peak position of the diffraction pattern $P(\kk)$~\cite{Gerbier2005,Kennedy2015}, which is defined as:
\begin{align}
  P(\kk)=\frac{1}{N_L^2}\sum_{\rr,\rr'}\rho_{\rr,\rr'}e^{i\kk\cdot(\rr-\rr')}.
\end{align}
For several degenerate lowest single-particle orbitals $[\kk_1,\cdots,\kk_q]$, take $P_c=\sum_{i=1}^{q}P(\kk_i)/q$. The reference background signal is the average of $P(\kk)$ over entire Brillouin zone $P_b=\sum_{\kk}P(\kk)/N_s$, and the condensed visibility is defined by the difference $P_c-P_b$. As shown in Fig.~\ref{phaseU}(c), at $U/t\ll1$, $P(\kk)$ has sharp peaks at momenta $[\kk_1,\cdots,\kk_q]$, while it vanishes at other momenta, and $q\times P_c$ is almost a constant. Otherwise, we also diagonalize the $N_{L}\times N_{L}$-matrix $\rho_{\rr,\rr'}$ and obtain one particle eigenstates $\rho|\phi_{\alpha}\rangle=n_{\alpha}|\phi_{\alpha}\rangle$ where $|\phi_{\alpha}\rangle$ ($\alpha=1,\ldots,N_{L}$) are the natural orbitals and $n_{\alpha}$ ($n_1\geq\ldots\geq n_{N_{L}}$) are interpreted as occupations. For $U/t\ll1$, we find that the occupations $n_{\alpha}\simeq N/q$ for $\alpha\leq q$, while $n_{\alpha}\ll1$ for $\alpha>q$, namely, a Bose condensate occurs~\cite{Mueller2006}. By increasing the interaction, $\rho_{\rr,\rr'}$ at $\max|\rr-\rr'|$ gradually decreases to a small value. For strong interaction $U\gg1$, $n_{\alpha}\simeq 1$ for $\alpha\leq N$, while $n_{\alpha}\ll1$ for $\alpha>N$. This is consistent with our observation that $|P_c-P_b|$ should be vanishingly small.

On the other hand, to evaluate its phase coherence, we impose a phase gradient $\theta$ through twisted boundary conditions~\cite{Melko2005}, and define the bosonic superfluid phase stiffness as
\begin{align}
  \rho_s=\lim_{\theta\rightarrow0}\frac{2}{N_L}\frac{E(\theta)-E(0)}{\theta^2}.
\end{align}
As illustrated in Fig.~\ref{phaseU}(c), for weak interaction $U<1$, $\rho_s$ has a finite large value indicating the superfluidity of the ground state, and begins to drop with the increase of $U$. Finally consistent with the diffraction peak, $\rho_s$ decreases to a small value for $U\gg1$. By finite-size scaling with  increasing particle numbers up to ten for $U=\infty$, exact diagonalization confirms that $\rho_s$ becomes vanishingly small.

In Fig.~\ref{phaseU}(d), the energy gap and many-body Chern number is plotted. For small $U\ll1$, $\Delta$ is zero. $\Delta$ is quite small for $U\sim1$, while it saturates to a value of the order of the band gap for strong interaction $U\gg1$. Meanwhile, the many-body Chern number of the ground state is quantized to $C_{mb}=2$ for $U\gtrsim1$ and $m\geq9$. However, for small $U$, its many-body Chern number is not well-defined due to the level crossing, and we do not plot it. For $U=0$, the many-body wave function for Bose condensate is a product of single particle orbits in the lowest $C=2$ band, $|\psi(\theta_x,\theta_y)\rangle=\sum_{\{\kk_{j}\}}\psi(\{\kk_{j}\})\prod_{j=1}^{N}\chi_{\kk_j}(\theta_x,\theta_y)$, where $\chi_{\kk_j}(\theta_x,\theta_y)$ is the single-particle Bloch state in the lowest band and $\kk_j \in [\kk_1,\cdots,\kk_q]$. Thus its many-body Chern number $C_{mb}=\frac{N}{q}\sum_{j=1}^{q}\oint d^2\theta \nabla_{\theta}\times(\chi_{\kk_j}^{\dag}\nabla_{\theta}\chi_{\kk_j})/2\pi i$, such that $C_{mb}$ usually does not exhibit a quantized behavior (for instance, $C_{mb}\simeq2.2$ for $N=N_y=5,m\geq24$) and may change with the interaction. In recent experiments, the non-quantized Hall response of a Bose condensate is observed in transport properties~\cite{LeBlanc2012} and charge pumping~\cite{Lu2015}. In order to identify the fluctuations in the crossover, we consider the one-particle occupation entropy $S=-\sum_{\alpha}n_{\alpha}\ln(n_{\alpha})$. In the inset of Fig.~\ref{phaseU}(d), we show the variance of the entropy as a function of interaction. In the limit $U=0$, the entropy approaches the negative value $-N\ln(N/q)$. In contrast, the entropy in the BIQH phase is a much smaller positive value.

\subsection{Fermi-liquid-like phase}

\begin{figure}[t]
  \includegraphics[height=2.8in,width=3.4in]{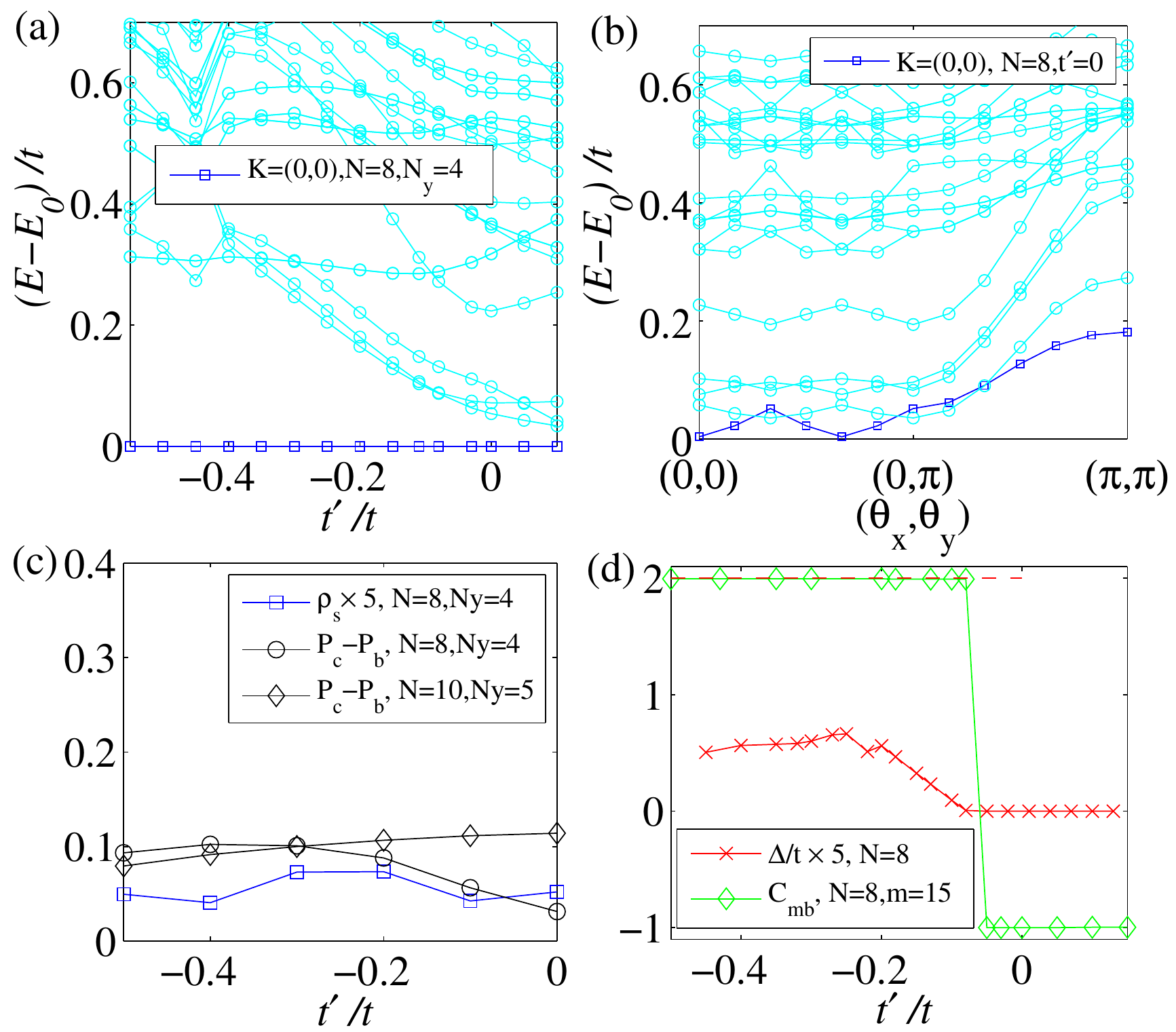}
  \caption{\label{phaset}(Color online) Numerical results for generalized Hofstadter model versus next nearest hopping $t'$ at infinite $U$: (a) the evolutions of the energy difference between the $K=(0,0)$ ground energy and the excited levels at $(\theta_x,\theta_y)=(0,0)$; (b) The low energy spectra flux under the insertion of $\theta_y$ at fixed $t'=0$ and $\theta_x=0$; (c) the diffraction peak of the visibility $P_c-P_b$ and the related superfluid phase stiffness $\rho_s$; (d) The energy gap $\Delta$ and many-body Chern number of the ground wavefunction at $K=(0,0)$ for $15\times15$ mesh points of Berry curvature. The red dashed line in (d) panel is the quantized value $C=2$.}
\end{figure}

\begin{figure}[t]
  \centering
  \includegraphics[height=1.7in,width=3.4in]{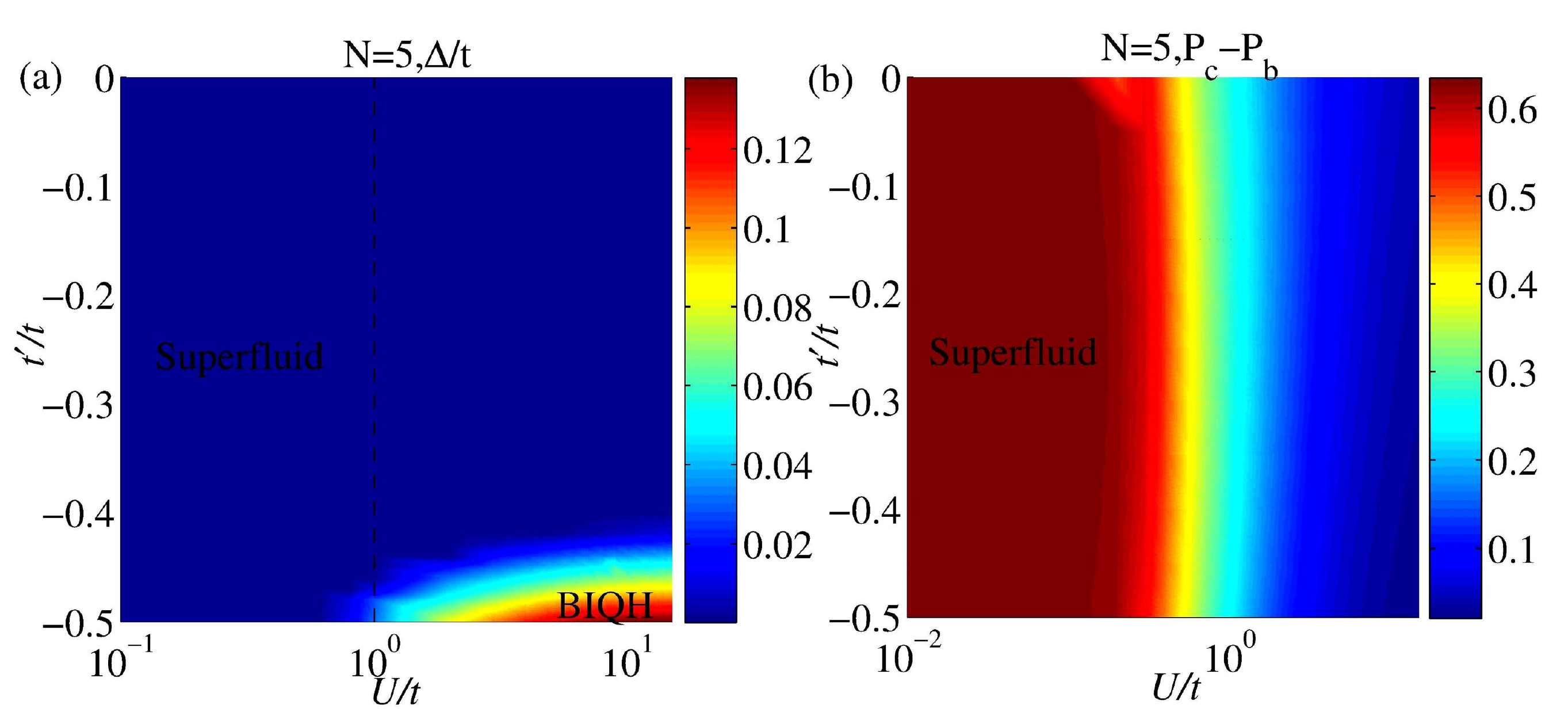}
  \caption{\label{phasediagram}(Color online) Numerical results for generalized Hofstadter model on the $(U,t')$-plane: (a) the intensity plot of the energy gap of $N=5,N_y=5$ system; (b) the diffraction peak of the visibility $P_c-P_b$ at $t'=0$.}
\end{figure}

Similarly, in Fig.~\ref{phaset}(a), we plot the evolution of the $K=(0,0)$ ground state for hardcore boson against next nearest hopping $t'$. For small $t'$ near $t'=0$, the $K=(0,0)$ ground state no longer maintain its robustness under the insertion of flux quantum, as indicated in Fig.~\ref{phaset}(b). Meanwhile, its superfluid stiffness and diffraction pattern show  featureless behavior during the transition, implying no possible superfluid phase. In Fig.~\ref{phaset}(d), one can see that the energy gap gradually drops to zero as $t'$ goes to zero, and the many-body Chern number changes rapidly for small $t'$ where the band topology structure is altered, and then drops to a negative value around $C_{mb}=-1$.

A typical picture of energy spectrum gap $\Delta$ in the $(U,t')$-plane is shown in Fig.~\ref{phasediagram}(a). The BIQH phase characterized by a finite gap is located at the right lower region where $U\gtrsim1,C=2$. The diffraction peak $P_c$ of the ground state at $(\theta_x,\theta_y)=(0,0)$ is shown in Fig.~\ref{phasediagram}(b). The left region where $U\lesssim1$ is characterized by a finite near constant condensation in single-particle orbits, regardless of the band topology, which is a superfluid phase.  The right region where $U\gg1$ implies no off-diagonal long range order. All these phases do not host any Bragg peak. Moreover, upon changing twisted boundary angles, except for BIQH, the ground state evolves into higher energy levels, indicating its metallic nature. For strongly interacting bosons in the right upper region where the lowest band topology changes to a band with $C=1$, the many-body Chern number of its ground state usually does not host integer quantized value once the disorder is introduced~\cite{Sheng2003}.

\begin{figure}[t]
  \centering
  \includegraphics[height=3in,width=3.4in]{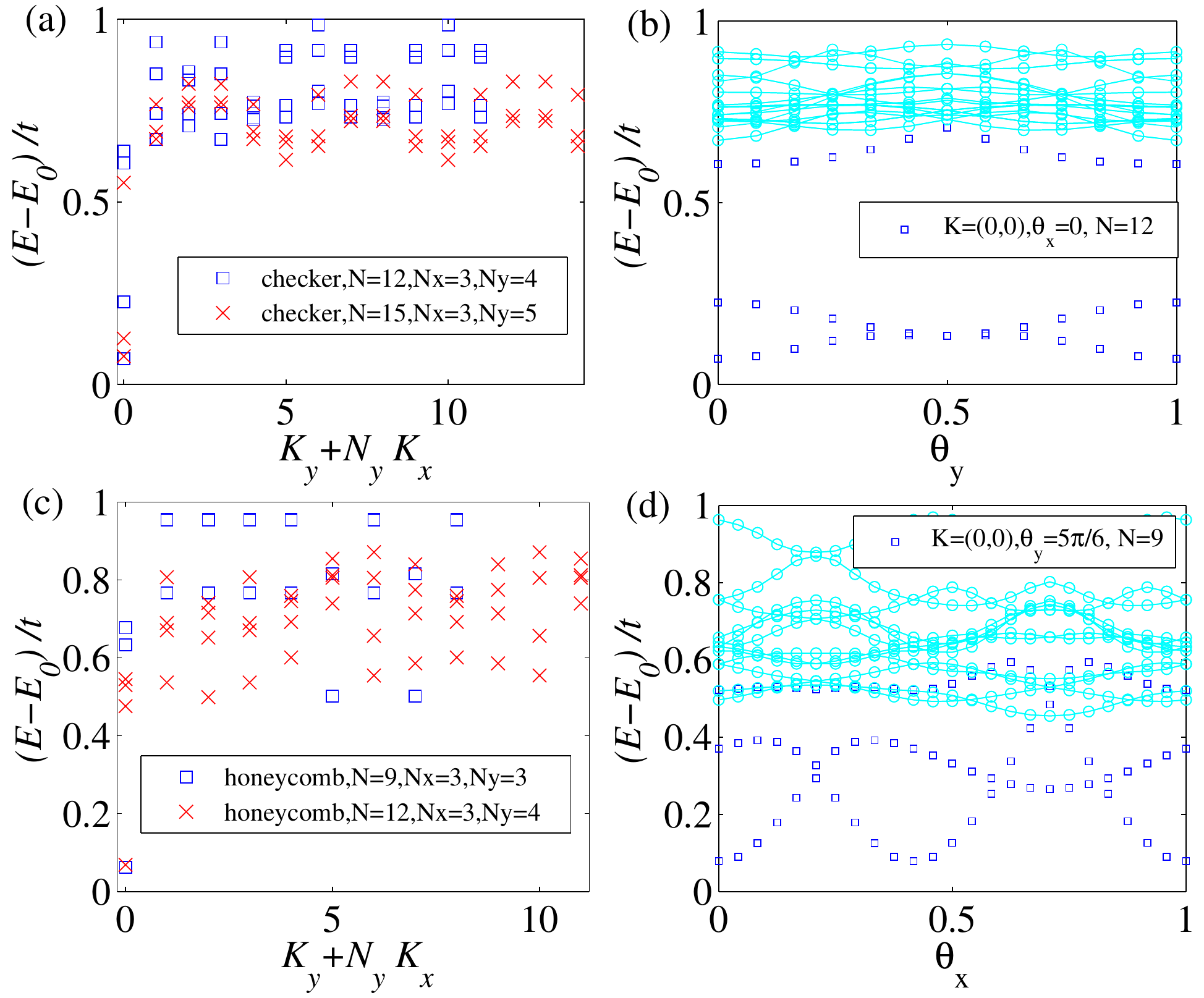}
  \caption{\label{metal}(Color online) Numerical results for hardcore bosons on two different topological lattices at filling $\nu=N/(N_xN_y)=1$ of the lowest Chern band $C=1$: (a) the low energy spectrum and (b) the low energy spectrum vs. flux  for checkerboard lattice; (c) the low energy spectrum and (d) the low energy spectrum vs.flux for honeycomb lattice. The checkerboard lattice hopping parameters are the same as Ref.~\cite{Zeng2015}, while the honeycomb lattice hopping parameters are the same as Refs.~\cite{Wang2011,Wang2012}}
\end{figure}

In comparison, we also present numerical results of hardcore bosons at $\nu=1$ on the topological checkerboard and honeycomb lattice models whose lowest band possesses topological invariant $C=1$~\cite{Sheng2011,Neupert2011,Tang2011,Sun2011,Regnault2011,Yao2013,Zeng2015,Wang2011,Wang2012}. As shown in Fig.~\ref{metal}, the low energy states in the same $K(0,0)$ sector evolve into each other under the insertion of flux quantum. We find that the Berry curvatures are vanishing small and the many-body Chern number of the ground state $C_{mb}\simeq0$. Both the intra-sublattice and inter-sublattice structure functions of density correlations
$S^{\alpha\beta}(\qq)=\sum_{\rr,\rr'}e^{i\qq\cdot(\rr-\rr')}n_{\rr}^{\alpha}n_{\rr'}^{\beta}/N_s$ have only a zero-momentum peak. One conjecture is that this phase would be a metallic Fermi-liquid like phase at $\nu=1$. We leave more details of this phase to be addressed in future studies.

\subsection{Charge density wave phase}

\begin{figure}[t]
  \includegraphics[height=3in,width=3.4in]{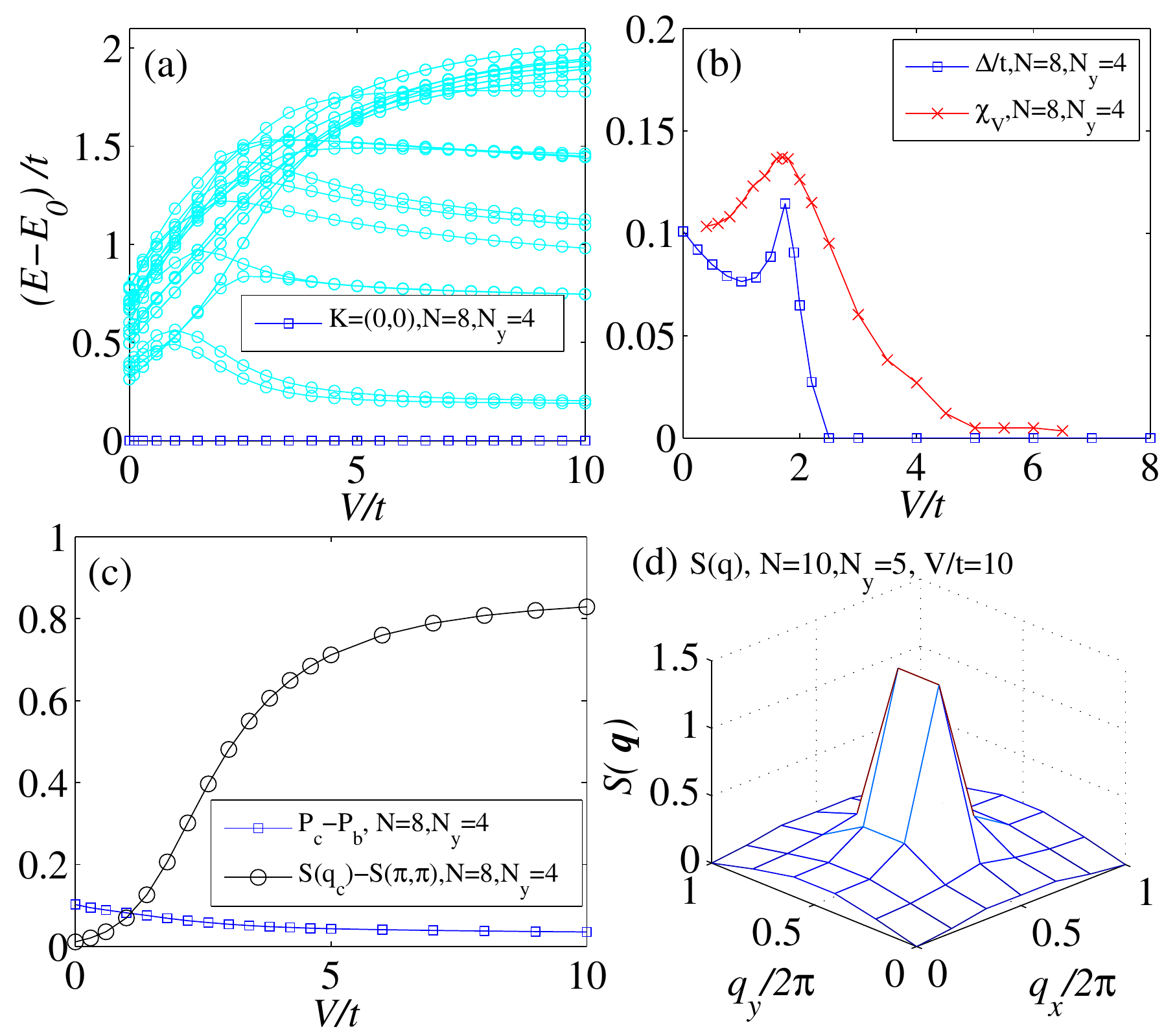}
  \caption{\label{phaseV}(Color online) Numerical results for generalized Hofstadter model at $U=\infty$,$(\theta_x,\theta_y)=(0,0)$ and $t'/t=-0.45$: (a) The evolutions of low energy states, (b) the energy gap $\Delta$ and fidelity susceptibility $\chi_V$ and (c) the peaks of density structure factor $S(\qq)$ and diffraction pattern versus nearest neighbor interaction $V$ for $N=8,N_x=2,N_y=4$; (d) The density structure factor at $V/t=10$ for $N=10,N_x=2,N_y=5$.}
\end{figure}

Finally, when the nearest neighbor repulsion is taken into account, we plot the variations of low energy states with $V$ in Fig.~\ref{phaseV}(a) from $V=0$ to $V\gg1$. The dependence of $\Delta$ on small $V$ is quite complicated, and it does not show a monotonic behavior. However for large $V$, the energy gap would collapse to zero indeed, as shown in Fig.~\ref{phaseV}(b). We calculate the fidelity susceptibility $\chi_V$ of the ground $K=(0,0)$ state, defined by
\begin{align}
  \chi_V=2\frac{1-|\langle\psi(V)|\psi(V+\delta)\rangle|}{\delta^2}.
\end{align}
$\chi_V$ exhibits a peak near the point where $\Delta$ collapses, which  serves as a signal of quantum phase transition~\cite{Gu2010}. In order to identify this transition, we plot the evolution of the peaks of its density structure factor and diffraction pattern in Fig.~\ref{phaseV}(c). For strong repulsion $V\gg t$, the peak of density structure factor $S(\qq)$ emerges at finite vector $\qq=\qq_c$, while the diffraction peak does not exhibit any upward jump behavior. By comparing  $S(\qq)$ for  larger sizes, e.g., exact diagonalization of up to ten particles shown in Fig.~\ref{phaseV}(d) and density matrix renormalization group of up to 18 particles, we confirm the existence of the Bragg peak. Thus the phase for $V\gg t$ is found  to be a charge density wave phase, instead of a superfluid or supersolid phase.

\section{Summary and discussion}\label{summary}

In summary, we have studied the Bose-Hubbard model in two dimensional generalized Hofstadter band with $C=2$, and demonstrated that strongly interacting bosons at filling $\nu=1$ can host BIQH state with Hall conductivity exactly quantized to 2 for strong onsite repulsion. The phase transition from bosonic superfluid to BIQH state driven by tuning the onsite Hubbard repulsive interaction is revealed by calculating superfluid phase stiffness and the diffraction pattern of the off-diagonal long range order. Tuning next nearest neighbor hopping down to zero changes the band topology, and the possible Fermi-liquid-like phase at at filling $\nu=1$ of $C=1$ Chern band is characterized by no phase coherence or well-defined many-body Chern number, gapless spectrum flux and featureless structure factors. Strong nearest neighbor repulsion would lead to a charge density wave.

In current $^{87}\text{Rb}$ experiments~\cite{Aidelsburger2013,Miyake2013,Kennedy2015}, magnetic flux $\phi$ can be tunable over the range $0\lesssim\phi\lesssim\pi$, onsite Hubbard interaction $U$ can be adjusted from zero up to $450$Hz by varying lattice potential depth and the nearest neighbor hopping $t\sim75$Hz, but the next nearest neighbor hopping is much smaller than nearest neighbor hopping. Experimental observations of these topological phases may be possible by further enhancement of next nearest neighbor hopping in the future, or using $\phi=4\pi/5$ with only small next nearest neighbor hopping where the lowest band also hosts Chern number two~\cite{Wang2013}. In our calculation, the energy gap $\Delta$ can be of the order of $t$, which is close to the cooling temperature limit. We believe that this Hofstadter model should provide a good platform for future study of topological phases.

\begin{acknowledgements}
This work is supported by the  U.S. Department of Energy,
Office of Basic Energy Sciences under Grant No. DE-FG02- 06ER46305. 

\end{acknowledgements}

\end{document}